\documentclass{elsart5p}
\usepackage{graphicx,epsfig,lineno}									
\usepackage{amssymb}
 \newcommand{\dqq}{\frac{\Delta Q}{Q}}

\begin{document}
\linenumbers
\begin{frontmatter}
\linenumbers

\title{A new approach in modeling the behavior of RPC detectors}

\author[lnf]{L. Benussi},
\author[lnf]{S. Bianco},
\author[lnf,sapienza,cern,COR]{S.Colafranceschi},
\author[lnf]{F.L. Fabbri},
%\author[lnf,sapienza]{F. Felli}
%\author[lnf,sapienza]{M. Ferrini}
\author[lnf]{M. Giardoni},
%\author[lnf,sapienza]{T. Greci}
%\author[lnf,sapienza]{A. Paolozzi}
\author[lnf]{L. Passamonti},
\author[lnf]{D. Piccolo},
\author[lnf]{D. Pierluigi},
\author[lnf]{A. Russo},
\author[lnf,sapienza]{G. Saviano},
%\author[bari,bari2]{M. Abbrescia}
\author[napoli1]{S. Buontempo},
\author[napoli1,napoli2]{A. Cimmino},
\author[napoli1,napoli2]{M. de Gruttola},
\author[napoli1]{F. Fabozzi},
\author[napoli1,napoli2]{A.O.M. Iorio},
\author[napoli1]{L. Lista},
\author[napoli1]{P. Paolucci},
%\author[bari]{AutoriBari}
\author[pavia]{P. Baesso},
\author[pavia]{G. Belli},
\author[pavia]{D. Pagano},
\author[pavia]{S.P. Ratti},
\author[pavia]{A. Vicini},
\author[pavia]{P. Vitulo},
\author[pavia]{C. Viviani},
%\author[cern]{AutoriCern}
\author[cern]{A. Sharma},
\author[cern]{A.~K.~Bhattacharyya}
\address[lnf]{INFN Laboratori Nazionali di Frascati, Via E. Fermi 40, I-00044 Frascati, Italy}
\address[sapienza]{Sapienza Universit\`a degli Studi di Roma
 \lq\lq La Sapienza\rq\rq, Piazzale A. Moro, Roma, Italy}
\address[cern]{CERN CH-1211 Gen\'eve 23 F-01631 Switzerland}
%\address[bari]{INFN Sezione di Bari, Via Amendola, 173I-70126 Bari, Italy}
%\address[bari2]{Dipartimento Interateneo di Fisica, Via Amendola, 173I-70126 Bar
%i, Italy}
\address[napoli1]{INFN Sezione di Napoli,
  Complesso Universitario di
Monte Sant'Angelo, edificio 6, 80126 Napoli, Italy }
\address[napoli2]{Universit\`a di Napoli Federico II, Complesso
Universitario di Monte Sant'Angelo, edificio 6, 80126 Napoli, Italy  }
\address[pavia]{INFN Sezione di Pavia and Universit\`a degli studi di
Pavia, Via Bassi 6, 27100 Pavia, Italy  }
\thanks[COR]{Corresponding author: Stefano Colafranceschi\\ E-mail address: 
stefano.colafranceschi@cern.ch}
  \begin{abstract}
  The behavior of RPC detectors is highly sensitive to environmental variables. 
A novel approach is presented to model the behavior of RPC detectors
 in a variety of experimental conditions. 
The algorithm, based on Artificial Neural Networks, has been developed 
and tested on the CMS RPC gas gain monitoring system during 
commissioning.  
  \end{abstract} 


\end{frontmatter}

%% \linenumbers

%% main text
%
\section{Introduction}
Resistive Plate Chamber (RPC) detectors \cite{Santonico:1981sc}
are widely used in HEP experiments for muon detection and triggering at high-energy,
high-luminosity hadron colliders \cite{cmscollab,atlascollab}, in astroparticle physics experiments for the
detection of extended air showers \cite{D'AliStaiti:2008zz}, as well as in medical and imaging applications \cite{imaging}. At
the LHC, the muon system of the CMS experiment\cite{:2008zzk} relies on drift tubes, cathode strip 
chambers and RPCs\cite{RPC}.  
\par
  In this paper a new approach is proposed to  model the behavior of an RPC detector via a
  multivariate strategy. Full details on the developed algorithm and results can be found in
  Ref.\cite{Benussi:2010NNCMSNOTE}.
  The  algorithm, based on Artificial Neural Networks (ANN),
   allows one to predict
  the behavior of RPCs as a function of a set of variables, once enough data is available
  to 
  provide a training to the ANN. At the present stage only environmental variables (temperature
  $T$, 
  atmospheric pressure $p$ and relative humidity $H$) have been considered. Further studies
  including radiation dose are underway and will be the subject of a forthcoming paper.
 In a preliminary phase we trained a neural network with just one variable and we 
found out, as expected, that the predictions are improved after
adding more variables into the network. The agreement found between data and prediction has to be
considered a pessimistic evaluation of the validity of the algorithm, since it also depends on
the presence of unknown variables not considered for training.
 \par
 The data for this study have been collected utilizing the 
 gas gain monitoring (GGM)
 system
\cite{Abbrescia:2007mu}\cite{Benussi:2008fp}\cite{Benussi:2008vs}
 of the CMS RPC muon detector during the commissioning with cosmic rays in the ISR test area at CERN.
%The GGM is described in details elsewhere
%\cite{Abbrescia:2007mu}\cite{Benussi:2008fp}\cite{Benussi:2008vs}.
 \par
The GGM system  is composed by the same type of RPC used in the CMS detector (2~mm-thick  Bakelite gaps) but of
smaller size (50$\times$50 cm$^2$). Twelve gaps are arranged in
a stack. The trigger is provided by four out of twelve
gaps of the stack, while the remaining eight gaps are used to monitor the working point by means of a  cosmic
ray telescope based on RPC detectors.
\par
 In this study, the GGM was
operated in open loop mode with a 
Freon 95.5\%, Isobutane 4.2\%, SF$_6$ 0.3\% gas mixture. Six out
of eight monitoring gaps were used, two out of eight monitoring gaps failed during the study and were therefore excluded from the analysis.
 The monitoring is performed by measuring
the charge distributions of each  chamber. The six gaps are operated at different high
voltages, fixed for each chamber, in order to monitor the total range of operating
modes of the gaps (Table ~\ref{TAB:HV}).
 The operation mode of the RPC changes as a function of
the voltage applied, in particular the chamber will change from avalanche mode to streamer mode when increasing HV.
 
\par
  \begin{table}[htb]
    \caption{Applied high voltage for power supplies for GGM RPC detectors used in this study}
    \label{TAB:HV}
    \begin{center}
      \begin{tabular}{|l|c|c|c|c|c|c|} \hline
               & CH1 & CH2 & CH3 & CH6 & CH7 & CH8 \\ \hline
 Applied high voltage (kV) & 10.2 & 9.8  & 10.0  & 10.4 & 10.2  &  10.4 \\ \hline
      \end{tabular}
    \end{center}
  \end{table} 
 \par

\section{The Artificial Neural Network simulation code} 
\label{SECT:ANN}
\begin{figure}[tbph]
  \begin{center}
    \resizebox{4cm}{!}{\includegraphics{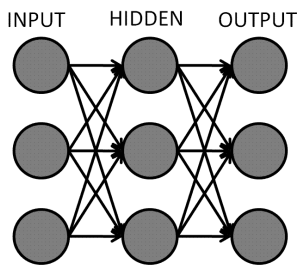}}
    \caption{Example of a simple Neural Network configuration.}
    \label{FIG:nnconf}
  \end{center}
\end{figure}

An Artificial Neural Network (ANN) is an information processing paradigm that is
inspired 
by the way biological nervous systems, such as the brain, process information\cite{MCCULLOCH}.
The most common type of artificial neural network (Fig.~\ref{FIG:nnconf}) consists of three groups, or
layers, of 
units: a layer of input units is connected to a layer of hidden units, which is
connected 
to a layer of output unit.
 The activity of the input units represents the raw information that is fed into
 the network.  
The activity of each hidden unit is determined by the activities of the input units
and 
the weights on the connections between the inputs and the hidden units.  
The behavior of the output units depends on the activity of the hidden units and
the 
weights between the hidden and output units.  
For this study temperature, humidity and pressure have been selected as inputs and
anodic 
charge as output variable. 
It was demonstrated\cite{hornik} that the number of layers is not critical for the
network performance, 
so  
we decided to go with 3 layers and give to the neural network a sufficient number of hidden
units 
automatically optimized by a genetic algorithm that can take into account several
configurations.  
\par
For each configuration a genetic algorithm 
performs the training process with an estimation of the global error; then  
the configuration is stored and the genetic algorithm continues to evaluate a slightly
different 
configuration. Once the algorithm has taken into account all the possible configurations 
the best one in terms of global error is chosen. 
\par
During the training phase the network is taught with environmental data as input, the
output depends on the 
neuronal weights, that at the very beginning are initialized with random numbers. 
The network output is compared to the experimental data we want to model, then the network
 estimates the error and modifies the neurons weights in order to minimize the estimated error.
\par
The training phase consists of determining both weights and configuration (nunber of neurons and number of layers)
 by minimizing the error, i.e., the difference between data and output.
\section{Environmental variables and datasets}
\label{SECT:DATASETS}
The environmental variables are monitored  by an Oregon Scientific
weather station WMR100.
The DAQ has been modified in order to acquire via USB the
environmental informations and merge environmental variables with
output variables.
The accuracy of the temperature sensor is $\pm 1^o$C
 in the range $0-40^o$C and
the resolution is $0.1^o$C. The relative humidity
sensor has an  operating range from 2\% to 98\%
 with a 1\% resolution, $\pm 7\%$ absolute accuracy from $25\%$ to
 $40\%$, and $\pm 5\%$  from $40\%$ to
 $80\%$.
The barometer operational range is  between 700~mbar and 1050~mbar
 with a 1~mbar resolution and a $\pm 10$~mbar accuracy.
\par
 The online monitoring system  records  the ambient
 temperature, pressure and humidity of the GGM box that contains the RPC stack.
 Pressure and temperature are mainly responsible of
 different detector behavior as well as the humidity for the bakelite and gas properties.

% ****BLUE BOX. WRITE TECHNICAL NOTE !
The used dataset is composed of four periods, each period composed of
 runs (about 270 each).
 Each run contains $10^4$ cosmic ray events where environmental
 variables and GGM  anodic output charges (\textit{Q}) are collected. 
The acquisition rate is typically 9.5~Hz.
\section{Results}
\label{SECT:RESULTS}
\begin{figure}[tbph]
  \begin{center}
    \resizebox{9.0cm}{!}{\includegraphics{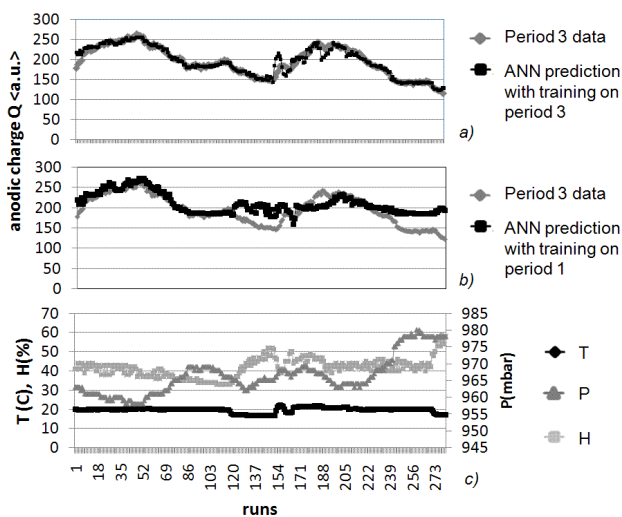}}
    \caption{ $(a)$ Gap 7 trained on the period 3 - prediction on period 3; the prediction is performed on the
 same period used as training with very good agreement between experimental data and prediction. 
$(b)$ Gap 7 trained on the period 1 - prediction on period 3, the prediction is performed on a period different from the training one, 
the agreement depends on dispersion of environmental variables.
$(c)$ Environmental variables during the period 3.}
    \label{FIG:CH7_OTT1_OTT1}
  \end{center}
\end{figure}

\begin{figure}[tbph]
  \begin{center}
    \resizebox{8.3cm}{!}{\includegraphics{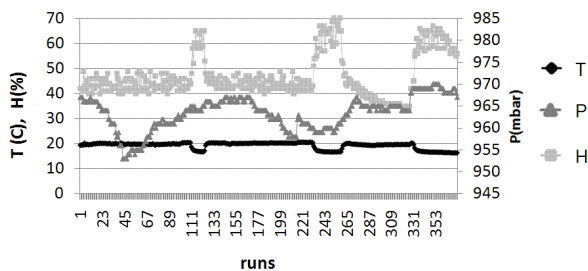}}
    \caption{Environmental variables during the period 1}
    \label{FIG:CH7_o1_s1}
  \end{center}
\end{figure}

Typical ANN outputs show generally good agreement between data and prediction during training phase.
(Fig.~\ref{FIG:CH7_OTT1_OTT1} $(a)$). In periods where the prediction is not accurate, the discrepancy is 
typically concentrated in narrow regions (\lq\lq spikes\rq\rq).
Fig~\ref{FIG:CH7_OTT1_OTT1} $(b)$ shows the prediction on period 3 using the period 1 as training, 
the discrepancy around run 137 and run 256 are  due to a set of environmental variables not available 
in the training period as shown in Fig.~\ref{FIG:CH7_OTT1_OTT1} $(c)$ and Fig.~\ref{FIG:CH7_o1_s1}.
\par
\begin{figure}[tbp]
  \begin{center}
%    \resizebox{10cm}{!}{\includegraphics{total_histo_trainNfore.eps}}
  \resizebox{6cm}{!}{\includegraphics{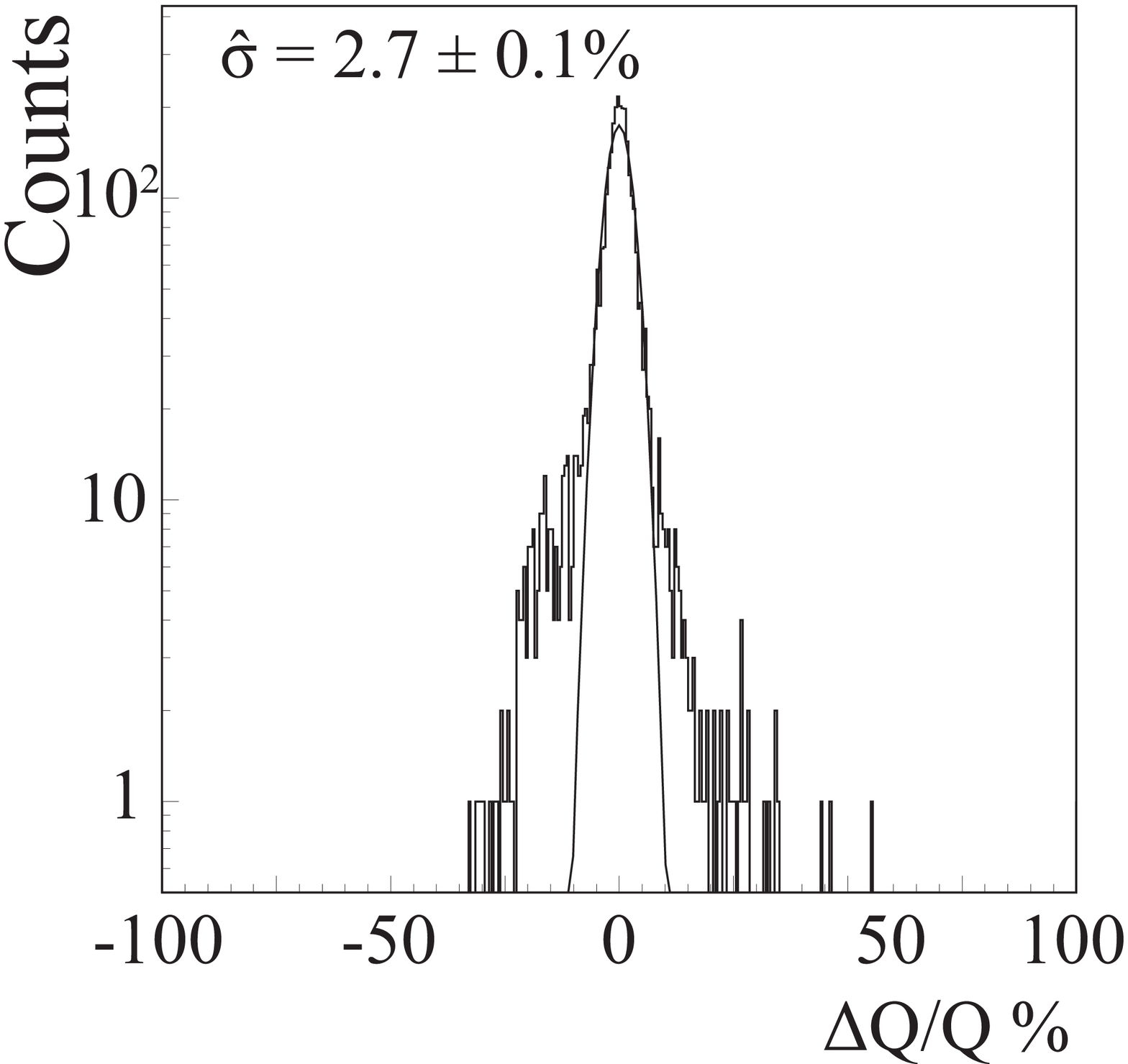}}
  \resizebox{6cm}{!}{\includegraphics{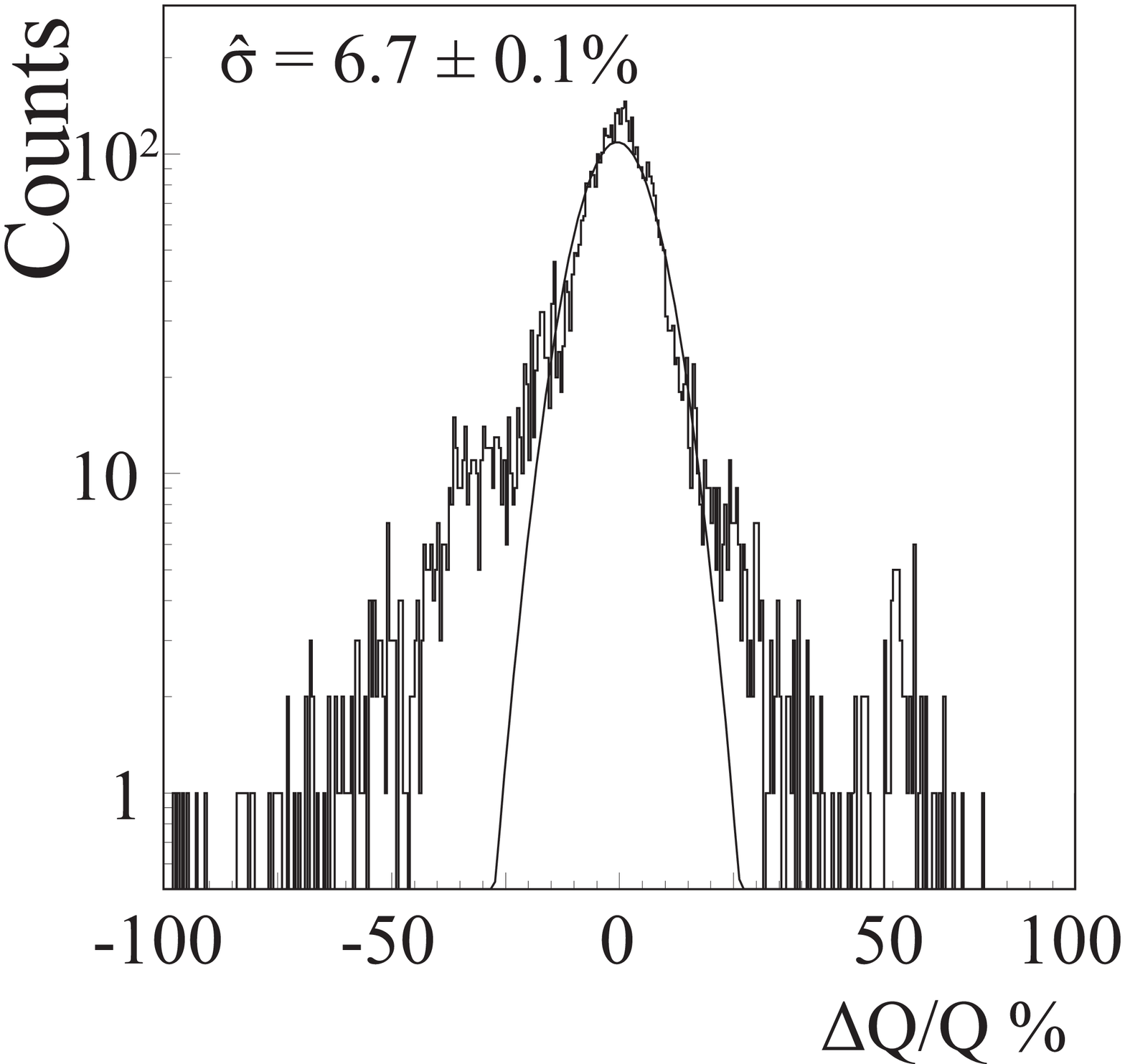}}
    \caption{Error for training (top) and
    prediction (bottom)  for all runs. Gaussian fit superimposed. The quantity
 $\hat {\sigma}$  is the width of the gaussian fit to the data in a reduced range which excludes the nongaussian tails.}
    \label{FIG:TOTHISTRAINFORE}
  \end{center}
\end{figure}

The comparison
between data and prediction is shown in
Fig.~\ref{FIG:TOTHISTRAINFORE} where the quantity 
\begin{equation}
 \dqq \equiv \frac{Q_{EXP}-Q_{ANN}}{Q_{EXP}}
\end{equation}
is plotted  for all four periods both for training (top) and predictions (bottom), divided for
training and prediction respectively. The error distribution for the predictions is much
wider than for the training, as expected.
\par
The gaussian fit superimposed (Fig.~\ref{FIG:TOTHISTRAINFORE}) is not able to fit the data properly due to the presence of large nongaussian tails, 
which are caused  by runs with very large discrepancy between data and prediction. To evaluate the width $\hat {\sigma}$ 
 of the error distribution we perform a gaussian fit in a reduced range which does not take into account the nongaussian tails.
 The distribution of the error for the predictions shows a $\hat{\sigma}$ = $6.7\%$.
 In the Table \ref{TAB:RESULTS} there is a summary with error for training and predictions.
 The cases with very large discrepancy were studied in detail, and found to
  be characterized by a $(p,T,H)$ value at the edges of the variables
   space. 
%  Fig.~\ref{FIG:PTH_VS_ERROR} shows the correlation between (p,T,H) and error.
%\par
%\begin{figure}[tbph]
% \begin{center}
%  \resizebox{7cm}{!}{\includegraphics{pressure_vs_error.eps}} 
%  \resizebox{7cm}{!}{\includegraphics{temp_vs_error.eps}} 
%  \resizebox{7cm}{!}{\includegraphics{rh_vs_error.eps}} 
%  \caption{(Top) pressure (center) temperature (bottom) humidity versus error 
 % $ \dqq $}
 % \label{FIG:PTH_VS_ERROR}
% \end{center}
%\end{figure}
\par
To determine the measure of the dispersion of the environmental variables considering all the runs ($N$) we computed the:
  \begin{equation}
      \frac{\Delta X}{X} \equiv \sqrt{\sum_{j=1,3} \biggr[\frac{(x_j-X_j)}{X_j}\biggr]^2  }
 \end{equation}

 \begin{equation}
    X_j \equiv \sum_{i=1,N} (x_j)_i \quad ; \quad {\bf x} \equiv (p,T,H) \quad 
 \end{equation}

\begin{figure}[tbph]
  \begin{center}
    \resizebox{8.5cm}{!}{\includegraphics{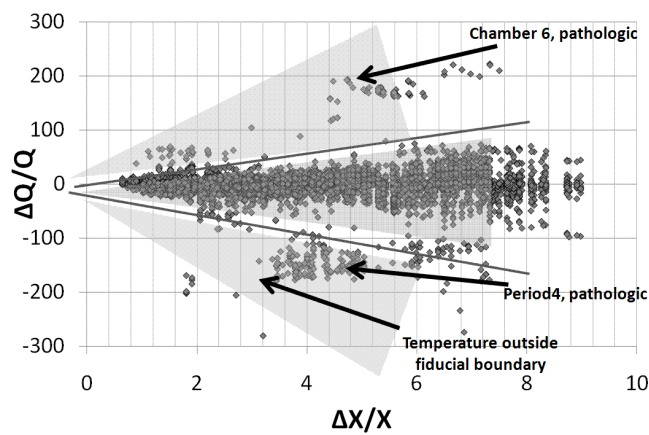}}
    \caption{Distribution of $\dqq$ as a function of the  dispersion of environmental variables $\frac{\Delta X}{X}$ for all periods,
    six gaps and both training and prediction. Each training period is included once, each prediction is included 4 times, due to different training period chosen.}
    \label{FIG:NORMA}
  \end{center}
\end{figure}

The distribution of the $\dqq$ error as a function of the dispersion of environmental variables $\frac{\Delta X}{X}$ 
 (Fig.~\ref{FIG:NORMA}) shows three distinct structures. The satellite bands with very large
 error were studied in detail. All data point in such bands belong to period four and gap
 six for which problems were detected. Period four and gap six therefore were excluded in
 the analysis. The distribution of the error as a function of dispersion of environmental variables
 after this selection has a
  $\hat {\sigma} \sim 4\%$ width and nongaussian tails extending up
  to $\dqq = 200\%$.
 \par
 A selection on the fiducial volume in the {\bf x} variables space 
 (Table ~\ref{TAB:SMARTCAZ}) was applied in order to exclude from the 
analysis data with $(p,T,H)$ close to the edges of the variable space.
After the selection cuts, prediction on two periods based on  training on the third period were performed. 
The nongaussian (NG) tails were defined as the fractional area outside the region $\pm 4 \hat{\sigma}$.
 The selection cuts slightly reduce the width ($\hat {\sigma} < 3.7\%$), while drastically reducing the nongaussian tails (Table \ref{TAB:RESULTS}).
 \par
  \begin{table}[htbh]
    \caption{Synopsis of the selection cuts for fiducial volume applied to predicted data.}
    \label{TAB:SMARTCAZ}
    \begin{center}
      \begin{tabular}{|c|c|c|} \hline
       $(958 < p < 968) {\rm mbar}$ & $ (19.4 <T< 20.4)^o$C  & $(34 < H < 44) \%$ \\ \hline
      \end{tabular}
    \end{center}
  \end{table}

 \par
  \begin{table}[htbh]
    \caption{Summary of errors $\hat {\sigma}$ and nongaussian (NG) tails for various selection cuts and samples. }
    \label{TAB:RESULTS}
    \begin{center}
      \begin{footnotesize}
      \begin{tabular}{|l|c|c|} \hline
   Data sets   &  $\hat {\sigma}$        & NG tail \\ 
               &     \%                  &       \%           \\ \hline
  
  All six chambers, all four periods   training                 & $2.7$  & $2.26$ \\  
  All six chambers, all four periods   prediction               & $6.7$  & $6.60$ \\ 
  Chamber six and period four excluded prediction               & $3.0$  & $4.63$ \\ 
  Predict. on per. 2 and 3, train. on per. 1                    & $4.0$ & $3.52$ \\ 
  Predict. on per. 3 and 1, train. on per.  2                 & $3.4$ & $2.95$ \\ 
  Predict. on per. 1 and 2, train. on per.  3                  & $3.8$ & $1.63$ \\ 
   Predict. on per. 2 and 3, train. on per. 1, fiducial cuts    & $3.7$ & $0.49$ \\ 
  Predict. on per. 3 and 1, train. on per.  2, fiducial cuts    & $2.9$ & $0.98$ \\ 
  Predict. on per. 1 and 2, train. on per.  3, fiducial cuts      & $3.3$ & $0.29$ \\ \hline
      \end{tabular}
      \end{footnotesize}
    \end{center}
  \end{table}

\section{Discussion}
In this study the GGM is the system used to train the neural network with anode charge as output variable and $(p,T,H)$ as input variable.
The addition of the dark current as a output variable and dose as input variable is expected to improve predictions and will be implemented.
The main advantage of this approach is that several variables can be used together in order to predict chamber behavior 
without the needs of studying the surface corrosion, environmental/radiation dependence and bakelite aging due to chemical reactions and deposits; 
also in the ANN analysis, given enough data, it is possible to decouple the effect of the chosen variables used as output.
This approach, once properly trained, 
could spot immediately and online pathological chambers whose behavior is shifting from the normal one.
 Further studies are in progress to determine and cure the residual nongaussian tails of the $\dqq$ errors
 distributions, 
to deal with training and prediction on detectors
with different high voltage supply, to widen the sample of environmental conditions, and in
adding new dimensions to the variables space such as radiation levels.

\section{Conclusions}
\label{SECT:CONC}
A new approach for modeling the RPC behavior, based on ANN, has been introduced and preliminary results obtained using data from the CMS RPC GGM system.
The ANN was trained for predicting the behavior of the anode charge Q (output variables) as function of the environmental variables $(p,T,H)$ (input variables),
 resulting in a prediction error $ \dqq = 4\%$. In a forthcoming work we plan to include the dose as input variable and the dark current as output variable, aiming
 at a further improvement on the predictions.
%
%figureTemplate
%\begin{figure}
%  \begin{center}
%    \resizebox{15cm}{!}{\includegraphics{closedLoop.eps}}
%    \caption{ fig.1 caption  placeholder for a figure
%     }
%    \label{FIG:SMLGAP}
%  \end{center}
%\end{figure}
%
%
\newline
\newline
{\bf Acknowledgements}
\newline
\newline
\label{SECT:ACK}
 The skills of M.~Giardoni, L.~Passamonti, D.~Pierluigi, B.~Ponzio and
 A.~Russo (Frascati) in setting up the experimental setup are
 gratefully acknowledged. The technical support of the CERN gas group
 is gratefully acknowledged.
  Thanks are due to R.~Guida (CERN Gas Group),
 Nadeesha M. Wickramage, Yasser Assran  for discussions and help.
 This research was supported in part by the Italian Istituto Nazionale
 di Fisica Nucleare and Ministero dell' Istruzione, Universit\`a e
 Ricerca.


\begin{thebibliography}{9}
\bibitem{Santonico:1981sc}
  R.~Santonico and R.~Cardarelli,
  %``Development Of Resistive Plate Counters,''
  Nucl.\ Instrum.\ Meth.\  {\bf 187} (1981) 377.
  %%CITATION = NUIMA,187,377;%%
%\cite{D'AliStaiti:2008zz}
\bibitem{cmscollab}
CMS Collaboration, %“The CMS experiment at the CERN LHC”,
 JINST 0803 (2008) S08004.
doi:10.1088/1748-0221/3/08/S08004.

\bibitem{atlascollab}
The ATLAS Collaboration, G. Aad et al., %The ATLAS Experiment at the
CERN Large Hadron Collider , JINST 3 (2008) S08003.

\bibitem{D'AliStaiti:2008zz}
  G.~D'Ali Staiti  [ARGO-YBJ Collaboration],
  %``The ARGO-YBJ experiment in Tibet,''
  Nucl.\ Instrum.\ Meth.\  A {\bf 588} (2008) 7.
  %%CITATION = NUIMA,A588,7;%%
%\cite{imaging}
\bibitem{imaging}
  P.~Fonte, %``Applications and New Developments in Resistive Plate Chambers``,
  IEEE Transactions on Nuclear Science, vol. 49, no. 3, June 2002.
\bibitem{:2008zzk}
  CMS Collaboration,
  %``The CMS experiment at the CERN LHC'',
  JINST {\bf 3} (2008) S08004.
  %%CITATION = JINST,3,S08004;%%
  \bibitem{RPC}
  CMS Collaboration,%``The CMS muon project: Technical Design Report``,
 CERN-LHCC-97-032 ; CMS-TDR-003. Geneva, CERN, 1997.
%\bibitem{ENVIRONM}
%  K.~Doroud et al. ``Simulation of resistive plate chamber in streamer mode operation``, Nucl. Instrum. and Meth. A {\bf 602} (2009)
%  723. 
%  M.~De~Vincenzi et al. ``Ageing and recovering of glass RPC, Nucl. Instrum. and Meth`` A {\bf 508} (2003)
%  94.
%  M.~Bianco et al. ``Resistive plate chamber commissioning and performance in CMS``, Nucl. Instrum. and Meth. A {\bf 602} (2009)
%  700.
%
\bibitem{Benussi:2010NNCMSNOTE}
 L.~Benussi et al., %``A new approach in modeling the behavior of RPC detectors''
 CMS NOTE 2010/076.
\bibitem{Abbrescia:2007mu}
  M.~Abbrescia {\it et al.},
  %``Gas analysis and monitoring systems for the RPC detector of CMS at LHC,''
 LNF-06-34-P, LNF-04-25-P, Jan 2007. 9pp. 
 Presented by S.~Bianco on behalf of the CMS RPC Collaboration at the
 2006 IEEE Nuclear Science Symposium 
 (NSS), Medical 
 Imaging Conference (MIC) and 15th International Room Temperature
 Semiconductor Detector Workshop, San Diego, California, 29 Oct - 4
 Nov  2006.  arXiv:physics/0701014.
  %%CITATION = PHYSICS/0701014;%%
%\cite{Benussi:2008fp}
\bibitem{Benussi:2008fp}
  L.~Benussi {\it et al.},
%``The CMS RPC gas gain monitoring system: an overview and preliminary results'',
 Nucl.\ Instrum.\ Meth.\  A {\bf 602} (2009) 805
  [arXiv:0812.1108 [physics.ins-det]].
  %%CITATION = NUIMA,A602,805;%%
%\cite{Benussi:2008vs}
\bibitem{Benussi:2008vs}
  L.~Benussi {\it et al.},
% ``Sensitivity and environmental behavior of the CMS RPC gas gain monitoring system,''
 JINST {\bf 4} (2009) P08006
  [arXiv:0812.1710 [physics.ins-det]].
  %%CITATION = JINST,4,P08006;%% 
\bibitem{MCCULLOCH}
   W.~S.~Mc~Culloch, W.~Pitts, % ``A logical  calculus of the ideas immanent in nervous activity'', 
Bulletin of Mathematical Biophysics {\bf 5} (1943) 
   115.  
\bibitem{hornik}
   K.~Hornik, M.~Stinchcombe and H.~White, %``Multilayer Feedforward Networks are Universal Approximators?'',
Neural Networks, vol. 2, pp.
    359, 1989.
\end{thebibliography}
\end{document}